# DYNAMICS OF A HIGH-DENSITY PLASMA IN A MAGNETIC FIELD

M M Tsventoukh

Lebedev Physical Institute of Russian Academy of Sciences, 53 Leninsky Ave., Moscow 119991, Russia

Abstract

The paper considers the dynamics of the plasma generated by the explosive-emission cells of the cathode spot of a vacuum arc in a magnetic field. It is shown that the expansion of the (high-density) plasma in a transverse magnetic field may cause asymmetry in the plasma density distribution at the cathode spot boundary. The asymmetry, in turn, increases the probability of the ignition of new explosive-emission cells in the region of a stronger magnetic field in the "anti-amperian" direction of **B**×**I**. The disturbed plasma density distribution estimated in the MHD approximation is presented. In addition, the velocity of the "retrograde" spot motion (ignition of new explosive-emission cells) in the stronger field region is estimated as a function of the external magnetic field strength. The velocity estimates (a few to tens of m/(s T)) are shown to agree with experimental data.

## 1. INTRODUCTION

Plasma of a large density in the magnetic field arises widely at powerful plasma sources and especially at fusion devices [1-3]. Plasma-surface interactions with the first wall of fusion devices can be accompanied by ignition of a self-sustained unipolar arcing producing dense plasma splashes, which is attributed to so-called vacuum electrical discharge [4-7]. An electrical discharge in vacuum involves production of plasma from the electrode material [8–11]. The current flowing through the plasma is due to the electron emission from the cathode accompanied by plasma production. It is commonly accepted that the electron emission and plasma production occur in a repetitively pulsed manner, much like boiling. The key features of a single pulse are conditioned by the event of explosive electron emission (EEE) occurring due to Joule overheating in a microvolume of the emission region within a few nanoseconds. The EEE-produced current pulse (called an ecton) is accompanied by the expansion of the metal plasma to the outside from the emission microregion at the cathode with velocities reaching 5–20 km/s. The ensemble of repetitively ignited EEE cells constitutes a cathode spot.

The successive events of ignition of new EEE cells and extinction of the preceding ones give an impression of a moving spot. When a vacuum arc operates in an applied magnetic field, the cathode spot exhibits a so-called "retrograde" motion, which is an average drift motion in the anti-amperian direction of **B**×**I** ($B$ is the magnetic field and $I$ is the arc current) [12, 13]. To describe this phenomenon, a variety of theoretical models have been proposed; however, it has not yet received a commonly accepted explanation. This paper considers the dynamics of the (dense) EEE plasma of a vacuum arc in a magnetic field with the aim to find out preferable conditions for the ignition of new EEE cells in the cathode spot.

In contrast to the early approaches, our model is based on the available results of experimental, theoretical, and computational investigations of the plasma of the explosive emission cells of a cathode spot operating in a repetitively pulsed manner [10].

It was found, in particular, that the initiation of an explosive emission pulse on a clean surface requires an excess of the threshold energy flux over the level of 100 MW/cm$^2$ [14]. This threshold can be achieved due to the energy brought by both the ionic and the electronic component of the plasma, depending on the sheath potential (see Refs. 14, 15).

In our opinion, the asymmetry in the probability density of the ignition of new EEE centers in a magnetic field is mainly caused by the expansion of the (high-density) explosive emission plasma [16]. Note that the ejection of plasma from the cathode of a vacuum discharge was considered as early as in 1904 by Weintraub [17] and in 1968 by Sena [18]. However, at that time, the parameters of the EEE plasma generated by cathode spot cells were not known, which made it impossible to develop a quantitative model.

## 2. GENERAL ESTIMATIONS

The dynamics of a plasma in a magnetic field can be characterized by dimensionless parameters, such as the plasma-to-magnetic pressure ratio $\beta = 8\pi nT/B^2$ and the magnetization parameter $\omega\tau$, where $\omega = eB/mc$ is the electron cyclotron frequency and $\tau = m^{1/2}T^{3/2}/(\pi e^4 \Lambda n)$ is the Coulomb collision time (where $\Lambda = \ln[(r_D^2 + \rho^2)^{1/2}/\rho]$ is the Coulomb logarithm, $r_D^2 = T/4\pi e^2 n$ is the Debye radius, and $\rho = e^2/T$).

We have derived general expressions of these parameters for the boundary of a plasma column of radius $R$ carrying a current of density

$$j = \kappa j_{pl} = \kappa e n \sqrt{\frac{T_e}{2\pi m_e}} \tag{1}$$

(where $\kappa < 1$ is the relative fraction of the electron current), which are written as [19]

$$\beta = \frac{4 m_e c^2}{e^2} \frac{1}{\kappa^2 n R^2} \tag{2}$$

$$\omega\tau = \sqrt{\frac{2}{\pi}} \frac{\kappa}{m_e c^2 e^2} \frac{T_e^2}{\Lambda} R. \tag{3}$$

Substitution of $m_e c^2$ = 0.511 MeV and $e^2$ = 0.144 µeV cm in Eqs. (2), (3) gives

$$\beta \kappa^2 n R^2 = 1.41 \times 10^{13} \text{ cm}^{-1}, \tag{4}$$

$$\omega\tau \frac{\Lambda}{\kappa T_e^2 R} = 10.84 \text{ eV}^{-2} \text{ cm}^{-1}. \tag{5}$$

These expressions indicate that for the parameter $\omega\tau$ at the boundary of an active EEE cell (where $n > 10^{18}$ cm$^{-3}$, $R < 10$ µm, $T_e \sim 1$ eV, and $\Lambda \sim 1$), we have

$$\omega\tau \ll 1 \ll \beta. \tag{6}$$

This implies that the self-magnetic field of an EEE cell has no influence on the dynamics of its plasma (although the field in this region is the strongest one). However, new cells may arise some distance away from the preceding ones. For this case, it is necessary to take account of the decrease in plasma density and in magnetic field with distance.

## 3. EQUATIONS AND PROBLEM GEOMETRY

For the case where the parameter $\omega\tau$ is not small, the MHD equations of momentum conservation for ions and electrons are written as [20]

$$Mn \frac{d\mathbf{v}}{dt} = -\nabla p + \frac{1}{c}[\mathbf{jB}], \tag{7}$$

$$\frac{\mathbf{j}}{\sigma} = \mathbf{E} + \frac{1}{c}[\mathbf{vB}] - \frac{1}{nec}[\mathbf{jB}],\quad (8)$$

where $M$ is the ionic mass, $n$ is the plasma density, $\mathbf{j} = en(\mathbf{v} - \mathbf{v_e})$ is the current density, $\sigma = ne^2\tau/m$, is the plasma conductivity, and rot$\mathbf{B} = 4\pi\mathbf{j}/c$. Consider the geometry of a cathode spot consisting of many cells.

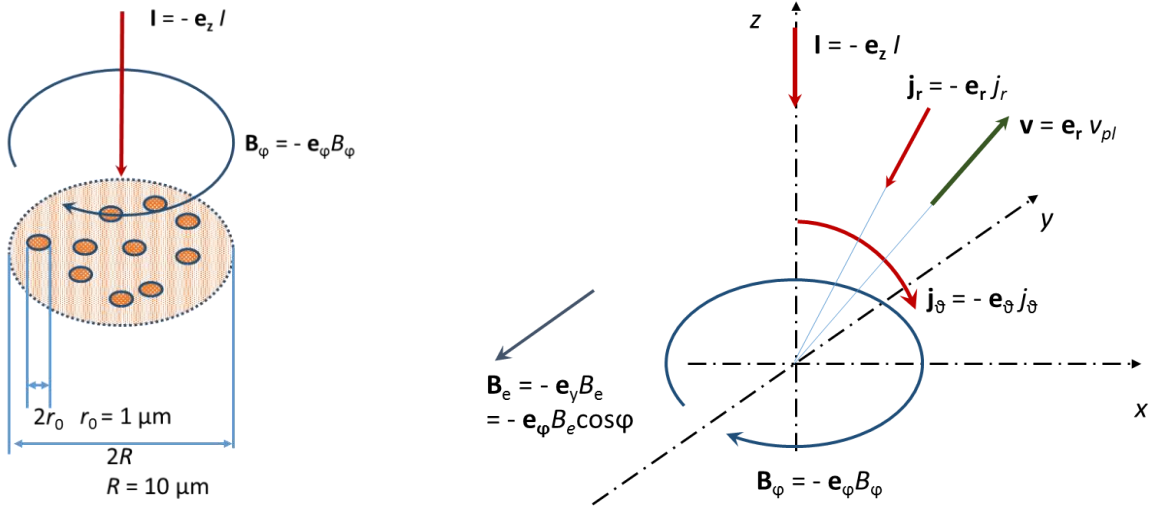

Fig. 1. Geometry of a cathode spot.

Ohm's law (8) can be written in components of spherical coordinates $(r, \vartheta)$ as

$$j_r = -Zenv_{1e} + \omega\tau j_\vartheta,\quad (9)$$

$$j_\vartheta = \omega\tau(Zenv_{pl} - j_r),\quad (10)$$

where $v_{1e}$ is the drift (current) velocity of electrons that provides the flow of the main current $I$ through the spot and $Z$ is the average ionic charge, which is over +1, +2, and +3 for Li, Cu, and W, respectively. According to our estimates, $v_{1e}$ averaged over an EEE pulse is on the level of the ion-sound velocity $(T/M)^{1/2} \sim 10^5$ cm/s [19]. This is comparable to the plasma expansion velocity $v_{pl}$, which is $\sim 10^5$ cm/s at a distance of 1 μm from the cathode spot and reaches $\sim 10^6$ cm/s at a distance of about 10 μm [21]. Thus, for the current $\mathbf{j}_\vartheta$ induced in the direction of $\mathbf{v} \times \mathbf{B}$, we have

$$j_\vartheta = \frac{\omega\tau}{1+\omega^2\tau^2} Zen(v_{pl} + v_{1e}).\quad (11)$$

Equation (7) indicates that this current may drive the amperian force reducing the acceleration of the plasma in the transverse magnetic field. The slower expansion of the plasma, according to the continuity equation, should result in an increase in its density. Note that the high density of the cathode spot plasma is the key factor in the ignition of new explosive-emission cells [14, 15, 22].

Consider the dependence of $\omega\tau$ on the distance from the cell. For $T \sim 1$ eV and $\Lambda \sim 1$, we have

$$\omega\tau = \frac{eB}{mc}\frac{\sqrt{m}T^{3/2}}{\pi\Lambda e^4 n} = \varpi\frac{B}{n} \approx 10^{13}\frac{B(r)}{n(r)}\ \text{cm}^{-3}\text{G}^{-1}.\quad (12)$$

It is well known that the cell plasma density $n_0$ at a distance $r_0 = 1$ μm is about $10^{20}$ cm$^{-3}$ on the average [9, 11, 23, 24]. The plasma density decreases with distance $r$ according to the relation $nr^2 \approx$ const or even more rapidly. Consider a relation of the form

$$n = n_0\left(\frac{r_0}{r}\right)^\alpha \quad (13)$$

with α = 2 and 3 for a more correct account of experimental data and calculation results [21, 25].

Assume that the magnetic field is created by a total current $I$ = 50 A flowing through a collective spot 10 μm in size (average current density $<j>$ = 16 MA/cm²) (see Ref. 26). Then relation (12) becomes

$$\omega\tau = \varpi \frac{2I}{cn_0 r_0^\alpha} r^{\alpha-1}. \tag{14}$$

For α = 2 and 3, we have two relations:

$$\omega\tau \big|_{1,\alpha=2} = \varpi \frac{2I}{cn_0 r_0^2} \times r \approx 10^2 r \text{ cm}^{-1}, \tag{15}$$

$$\omega\tau \big|_{1,\alpha=3} = \varpi \frac{2I}{cn_0 r_0^3} \times r^2 \approx 10^6 r^2 \text{ cm}^{-2}, \tag{16}$$

which indicate that ωτ increases with distance and reaches unity at a distance of about 100 and 10 μm for a quadratic and a cubic decrease in density, respectively.

As our estimate of the drift (current) velocity of electrons is comparable to the ion-sound velocity, the ion current may contribute to the total magnetic field of the spot [19]. Consider the total current at a distance $r > R$ for a quadratic and a cubic decrease in plasma density, respectively:

$$I_{r>R} = I \times [1 + (\zeta + \tfrac{1}{\kappa}\sqrt{\tfrac{2\pi m_e}{M_i}}) 2\ln\tfrac{r}{R}], \tag{17}$$

$$I_{r>R} = I \times [1 + (\zeta + \tfrac{1}{\kappa}\sqrt{\tfrac{2\pi m_e}{M_i}}) 2r_0(R^{-1} - r^{-1})], \tag{18}$$

where ζ is the fraction of the electron current emitted from the surface away from the spot. Assume that ζ = 0; hence, the electron current fraction is κ = (2πm/M)^(1/2) (see Eq. (1)). Then the specified relations for ωτ as a function of distance become

$$\omega\tau = \frac{e}{mc^2} \frac{\sqrt{m}T^{3/2}}{\pi\Lambda e^4} \frac{2I}{n_0 r_0^2} \times r[1 + 2\ln\tfrac{r}{R}] \approx 10^2 \times r[1 + 2\ln\tfrac{r}{10\mu m}] \text{ cm}^{-1}. \tag{19}$$

$$\omega\tau = \frac{e}{mc^2} \frac{\sqrt{m}T^{3/2}}{\pi\Lambda e^4} \frac{2I}{n_0 r_0^3} \times r^2[1 + 2r_0(\tfrac{1}{R} - \tfrac{1}{r})] \approx 10^6 \times r^2[1 + 2\,\mu m(\tfrac{1}{10\mu m} - \tfrac{1}{r})] \text{ cm}^{-2}. \tag{20}$$

In the first case, the distance corresponding to the transition of ωτ through unity decreases to 30 μm, and in the second case, it is 10 μm (which agrees with the estimate obtained from (16)).

Thus, it can be expected that at a distance of about 10 μm, which is comparable to the size of the collective spot, $R$, the effect of the self-magnetic field will be strong enough to induce a current of density $j_\vartheta$ (see Eq. (11)) in the direction of $v_{pl} \times B$.

It is worth noting that the plasma density (13) corresponding to this distance is about $10^{17} - 10^{18}$ cm$^{-3}$ that is close to the necessary EEE-cell ignition threshold, see Refs. 14-15.

4. INFLUENCE OF THE SELF-MAGNETIC FIELD

Estimate the effect of the induced current in view of the obtained relations for ωτ (Eqs. (15), (16)). The equation of motion for the radial velocity of the plasma expansion can be written as

$$M\frac{dv}{dt} = Mv\frac{dv}{dr} \approx -T\nabla \ln n - \frac{1}{c}\frac{\omega\tau}{1+\omega^2\tau^2} Ze(v + v_{1e})B_\phi. \tag{21}$$

The magnetic field $B_1$ generated by the current $j_\vartheta$ can be estimated from the relation $rot\mathbf{B}_1 = 4\pi \mathbf{e}_\vartheta j_\vartheta/c$.

Consider the case $\omega\tau <= 1$; then we have $B_1 = 4\pi/c \int j_\vartheta dr \approx 4\pi/c\ Zev_{pl} \int n\omega\tau dr \approx 100$ G, which is much lower than the field $10^4$ G created by the spot current at a distance of 10 μm. Let us take into account that the Ampere force – second term on the right hand side of (21), is much less than gradient acceleration term. Thus, we obtain

$$Mv\frac{dv}{dr} \approx -T\frac{d}{dr}\ln\frac{n}{n_0} - 2I\frac{Ze(v+v_{1e})}{c^2}\frac{\omega\tau}{r}, \tag{22}$$

and after integration over radius $r$ from $r_0$ to $R$ we obtain for the plasma velocity

$$v = \sqrt{\frac{2}{M}[\alpha T \ln\frac{R}{r_0} - \frac{Zev_{1e}}{c}\varpi\frac{R^{\alpha-1}}{n_0 r_0^\alpha}\frac{4I^2}{(\alpha-1)c^2}]} - \frac{Ze}{Mc}\varpi\frac{R^{\alpha-1}}{n_0 r_0^\alpha}\frac{4I^2}{(\alpha-1)c^2}. \tag{23}$$

The first term on the right side of relation (23), which describes the hydrodynamic acceleration of the plasma, can be estimated to be 4.6 and 7 eV, whereas the second one to be 2 and 10 meV for $\alpha = 2$ and 3, respectively. Write down the continuity equation in its simple form of a flux conservation law:

$$nv = const = n_{00}\sqrt{\frac{2}{M}(-T\ln\frac{n}{n_0})} = n_{00}\sqrt{\frac{2}{M}\alpha T \ln\frac{R}{r_0}}, \tag{24}$$

where $n_{00}$ is the plasma density obtained with no account of the magnetic field effect. Then we can estimate the plasma density increasing due to the confinement of the plasma by the current $j_\vartheta$ as

$$1 + \frac{\delta n}{n_{00}} = 1 \bigg/ (\sqrt{1 - \frac{1}{\alpha T \ln R/r_0}\frac{Zev_{1e}}{c}\varpi\frac{R^{\alpha-1}}{n_0 r_0^\alpha}\frac{4I^2}{(\alpha-1)c^2}} - \frac{1}{\sqrt{2M\alpha T \ln R/r_0}}\frac{Ze}{c}\varpi\frac{R^{\alpha-1}}{n_0 r_0^\alpha}\frac{4I^2}{(\alpha-1)c^2}) \approx$$

$$\approx 1 + (\frac{v_{1e}}{2\alpha T \ln R/r_0} + \frac{1}{\sqrt{2M\alpha T \ln R/r_0}})\frac{Ze}{c}\varpi\frac{R^{\alpha-1}}{n_0 r_0^\alpha}\frac{4I^2}{(\alpha-1)c^2}. \tag{25}$$

Thus, the increase in plasma density in the self-magnetic field can be estimated as $\delta n/n \sim 10^{-3}$.

5. INFLUENCE OF THE EXTERNAL MAGNETIC FIELD

Consider the effect of the external magnetic field directed along the cathode surface, $\mathbf{B}_e = -\mathbf{e}_y B_e = -\mathbf{e}_\varphi B_e \cos\varphi$ (see Fig. 1). Relation (14) describing $\omega\tau$ as a function of $r$ becomes

$$\omega\tau = \varpi\frac{r^\alpha}{n_0 r_0^\alpha}(\frac{2I}{cr} + B_e\cos\varphi). \tag{26}$$

For $\alpha = 2$ and 3, and $I = 50$ A, we then have

$$\omega\tau\big|_{2,\alpha=2;3} = \omega\tau\big|_{1,\alpha=2;3} \times (1 + \frac{crB_e\cos\varphi}{2I}) \approx \omega\tau\big|_{1,\alpha=2;3} \times (1 + \frac{rB_e\cos\varphi}{10\ cm\ G}), \tag{27}$$

(see Eqs. (15) and (16) for $\omega\tau|_{1,\alpha=2}$ and $\omega\tau|_{1,\alpha=3}$).

The angle $\varphi = 0$ corresponds to the retrograde direction of $\mathbf{B}\times\mathbf{I}$ and the angle $\varphi = \pi$ – to the amperian direction of $\mathbf{I}\times\mathbf{B}$. Putting $\omega\tau$ equal to unity, we obtain relations between the external field strength $B_e$ and the distance $r$ ($\omega\tau = 1$). The solid curves in Fig. 2 correspond to the "retrograde" side of the field, **B×I**, the dash-dot curves to the "amperian" side, **I×B**, and the dashed curves to $\omega\tau = -1$ on the "amperian" side, **I×B**. The last case corresponds to a strong external magnetic field (compared with the self-magnetic field of the spot) at which the induced current $\mathbf{j}_\vartheta$ changes its direction. (The case of a low cell current is considered in [16].)

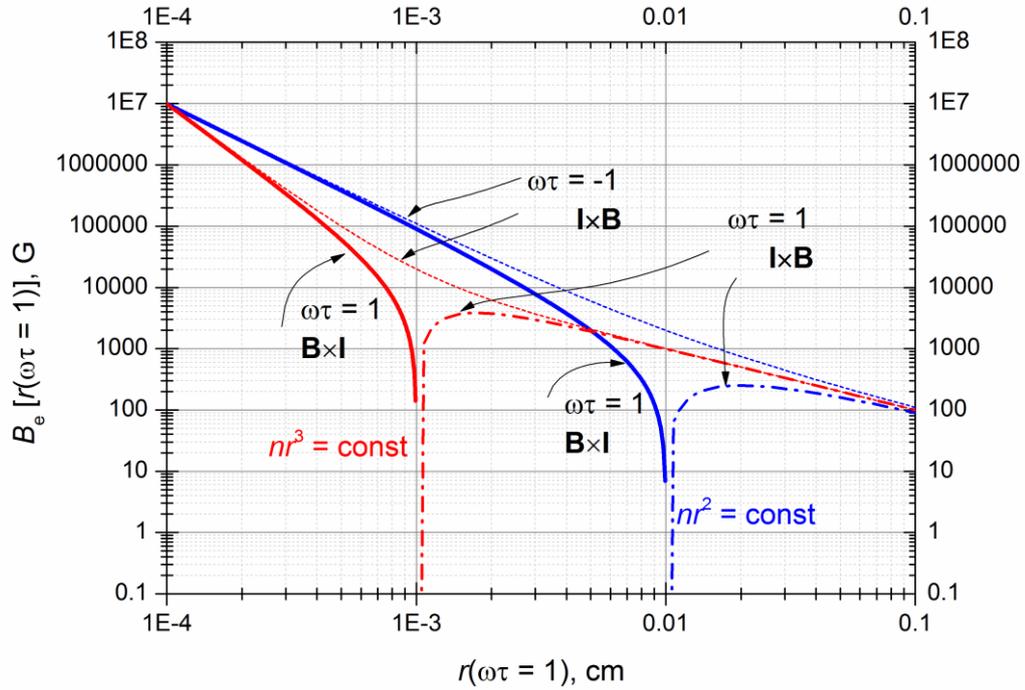

Fig. 2. The external magnetic field versus the distance from the center of an EEE cell corresponding to $|\omega\tau| = 1$ (Eq. (27)) for the undisturbed plasma density distributions (Eq. (13)) $nr^2$ = const and $nr^3$ = const on the retrograde and the amperian side.

As can be seen from the plots in Fig. 2, on application of an external magnetic field, the electronic component of the current becomes magnetized ($\omega\tau = 1$) nearer to the center of the cell on the retrograde side and father from the center on the amperian side. As the plasma density decreases with distance, the density of the plasma involved in deceleration by the induced current $j_\vartheta$ will be higher on the retrograde side.

The equation that describes the motion of the plasma in an external magnetic field can be written as

$$M\frac{dv}{dt} = Mv\frac{dv}{dr} \approx -T\nabla \ln n - \frac{1}{c}\frac{\omega\tau}{1+\omega^2\tau^2} Ze(v + v_{1e})(\frac{2I}{cr} + B_e\cos\varphi) \ . \tag{28}$$

Assume, as previously, that $\omega\tau <= 1$; then for the plasma velocity we have

$$v = \sqrt{\frac{2}{M}(\alpha T \ln\frac{R}{r_0} - v_{1e}\Upsilon)} - \frac{\Upsilon}{M}, \tag{29}$$

with function $\Upsilon$

$$\Upsilon \equiv \frac{Ze}{c}\varpi \frac{R^{\alpha-1}}{n_0 r_0^\alpha}[\frac{4I^2}{(\alpha-1)c^2} + \frac{4IB_e\cos\varphi}{\alpha c}R + \frac{B_e^2\cos^2\varphi}{\alpha+1}R^2]. \tag{30}$$

Using the continuity equation (24), we can obtain the following corrections to the plasma density, δn/n (see Fig. 3):

$$1+\frac{\delta n}{n_{00}} = 1\bigg/\left(\sqrt{1-\frac{v_{1e}\Upsilon}{\alpha T \ln R/r_0}} - \frac{\Upsilon}{\sqrt{2M\alpha T \ln R/r_0}}\right) \approx 1+\left(\frac{v_{1e}}{2\alpha T \ln R/r_0} + \frac{1}{\sqrt{2M\alpha T \ln R/r_0}}\right)\Upsilon. \quad (31)$$

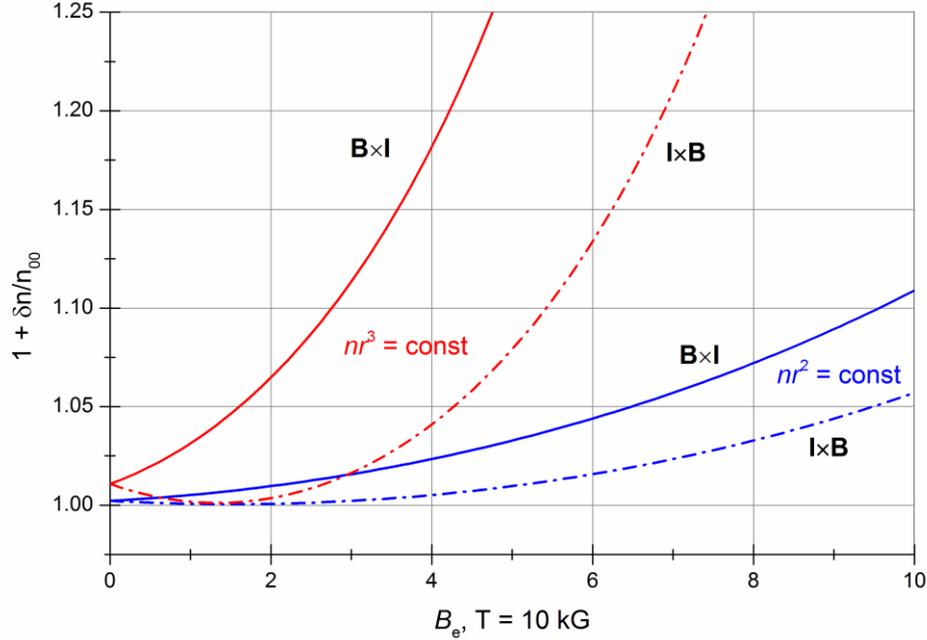

Fig. 3. Variations in density of the EEE plasma decelerated by the induced current (Eq. (11)) in a transverse magnetic field at ωτ ~ 1 (~10 μm away from the EEE center). The current *I* = 50 A

The velocity of the apparent spot motion can be estimated as the spot size *R* divided by the characteristic time of its ignition and operation, τ_ign (tens of nanoseconds). Assume that the plasma density stands for the probability of the occurrence of new emission centers (according to the results of the relevant investigations [14, 15, 22]). Then the average velocity of motion of the centers can be estimated as

$$<v>\big|_{\mathbf{B}\times\mathbf{I}} \equiv v_{retr} = \frac{R}{\tau_{ign}} \times \frac{n\big|_{\mathbf{B}\times\mathbf{I}} - n\big|_{\mathbf{I}\times\mathbf{B}}}{n}, \quad (32)$$

where the disturbed plasma density distribution is described by relation (31) for φ = 0 (corresponding to **B×I**) and φ = π (corresponding to **I×B**) (Fig. 4).
Also, in view of the expansion (31), we can obtain an analytic expression for the velocity of the retrograde motion of a cathode spot:

$$v_{retr}\frac{\tau_{ign}}{R} = \frac{\delta n\big|_{\varphi=0} - \delta n\big|_{\varphi=\pi}}{n_{00}} \approx \left(\frac{v_{1e}}{2\alpha T \ln R/r_0} + \frac{1}{\sqrt{2M\alpha T \ln R/r_0}}\right)\times\left(\Upsilon\big|_{\varphi=0} - \Upsilon\big|_{\varphi=\pi}\right) =$$

$$= \left(\frac{v_{1e}}{2\alpha T \ln R/r_0} + \frac{1}{\sqrt{2M\alpha T \ln R/r_0}}\right)\frac{Ze}{c}\varpi\frac{R^\alpha}{n_0 r_0^\alpha}\frac{8}{\alpha c} I B_e \quad (33)$$

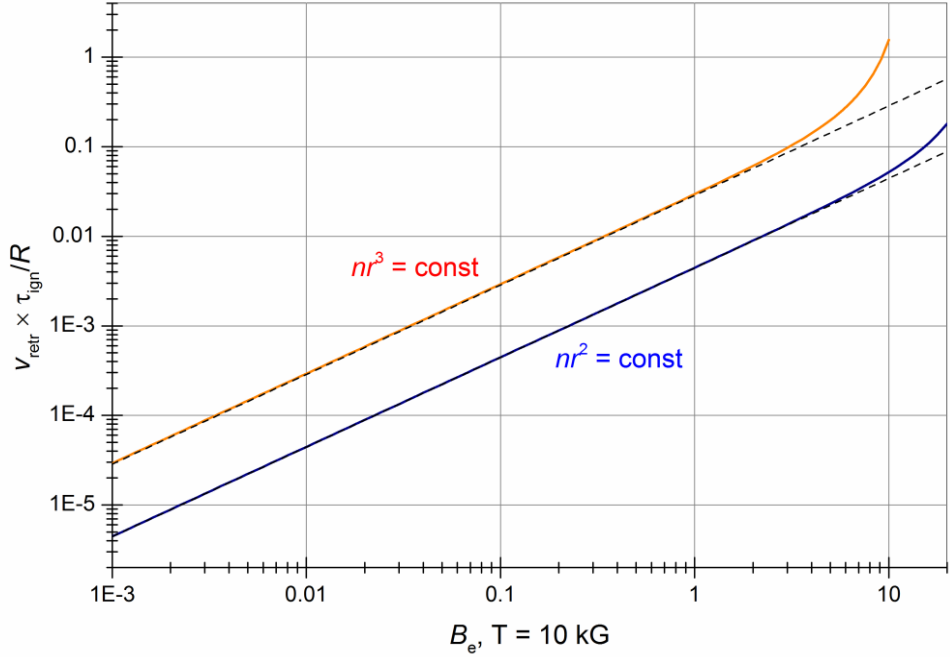

Fig. 4. The average velocity of the retrograde motion of a cathode spot (50-A spot current). The dashed lines are plots of the expansions.

The velocity of the "chaotic motion" of a cathode spot (occurrence of new cells), $R/\tau_{ign}$, is in the range $10^4$–$10^5$ cm/s; thus, we obtain the following relations between the velocity and the magnetic field:

$$\left.\frac{v_{retr}}{B_e}\right|_{\alpha=2} \approx 0.4 - 4.4 \ \frac{\text{m}}{\text{s T}}, \quad (34)$$

$$\left.\frac{v_{retr}}{B_e}\right|_{\alpha=3} \approx 2.9 - 28.6 \ \frac{\text{m}}{\text{s T}} \quad (35)$$

for $nr^2$ = const and $nr^3$ = const, respectively.

The estimated values of $v_{retr}/B_e$ ranging from a few to tens of m/(s T) generally agree with measurements (see Refs. 26, 27).

One may find that Eq. (33) predicts increasing in the average velocity of the spot retrograde motion with both current and magnetic field and simple estimation can be written as

$$v_{retr} \sim \frac{R}{\tau_{ign}} \times 10^{\alpha-6} \ I \ B_e \ \text{T}^{-1}\text{A}^{-1}. \quad (36)$$

6. DISCUSSION

Note that we have considered model (quadratic and cubic) relations for a decreasing plasma density. A more accurate consideration requires direct investigations of the plasma–surface interaction near active cells using both numerical simulation and experimental methods (laser interferometry [28]).

Also note that we have considered, for simplicity, the boundary case of ωτ approaching unity in some region. It seems that a more correct description of the phenomenon under consideration should involve larger spatial regions at the periphery of the cathode spot where the electronic component of the emitted plasma is magnetized more substantially.

For the cases of no and very strong magnetic field, relation (30) for the plasma velocity becomes, respectively,

$$v \propto \sqrt{1 - a_1 I^2} \ , \tag{37}$$

$$v \propto \sqrt{1 - a_2 B_e^2 \cos^2 \varphi} \ , \tag{38}$$

and relation (31) for the variations in plasma velocity it becomes, respectively,

$$1 + \frac{\delta n}{n_{00}} = 1 \Big/ \sqrt{1 - a_1 I^2} \approx 1 + \tfrac{a_1}{2} I^2 \ , \tag{39}$$

$$1 + \frac{\delta n}{n_{00}} = 1 \Big/ \sqrt{1 - a_2 B_e^2 \cos^2 \varphi} \approx 1 + \tfrac{a_2}{2} B_e^2 \cos^2 \varphi \ . \tag{40}$$

This corresponds, in the case $B_e = 0$ (considered in the beginning of the paper), to a symmetric compression of the plasma throughout the angle φ and, in the case of a strong external magnetic field, to a compression along the $x$-axis aligned with the direction of **B×I**.

7. CONCLUSIONS

We have considered the plasma generated by the explosive-emission cells of the cathode spot of a vacuum arc in a magnetic field. The plasma expanding in a transverse magnetic field has been shown to cause asymmetry in the plasma density distribution at the spot boundary. This increases the probability that new explosive-emission cells will be ignited in the region of a stronger magnetic field, namely, in the anti-amperian direction of **B×I**. For the disturbed plasma density distribution, estimates have been obtained in the MHD approximation. The velocity of the retrograde motion of the cathode spot, which appears as a result of occurrence of new explosive-emission cells in the region of a stronger magnetic field, has been estimated to range, depending on the magnetic field strength, from a few to tens of m/(s T), which agrees with experimental data.

Work was supported in part by RFBR grant # 16-08-01306.